\documentclass[a4paper,10pt]{article}

\usepackage{amsmath}
\usepackage{amsfonts}
\usepackage{amssymb}

\title{Metric fluctuations and decoherence}

\author{Heinz-Peter Breuer$^{1,2}$, Ertan G\"okl\"u$^3$ and Claus L\"ammerzahl$^3$ \\
$^1$ Physikalisches Institut, Universit\"at Freiburg,
Hermann-Herder-Strasse 3,\\
79104 Freiburg, Germany \\
$^2$ Hanse--Wissenschaftskolleg, Institute for Advanced Study, \\
27753 Delmenhorst, Germany \\
$^3$ ZARM, University of Bremen, Am Fallturm, 28359 Bremen,
Germany}

\date{\today}

\begin{document}

\maketitle

\begin{abstract}
Recently a model of metric fluctuations has been proposed which
yields an effective Schr\"o\-din\-ger equation for a quantum
particle with a modified inertial mass, leading to a violation of
the weak equivalence principle. The renormalization of the
inertial mass tensor results from a local space average over the
fluctuations of the metric over a fixed background metric. Here,
we demonstrate that the metric fluctuations of this model lead to
a further physical effect, namely to an effective decoherence of
the quantum particle. We derive a quantum master equation for the
particle's density matrix, discuss in detail its dissipation and
decoherence properties, and estimate the corresponding decoherence
time scales. By contrast to other models discussed in the
literature, in the present approach the metric fluctuations give
rise to a decay of the coherences in the energy representation,
i.~e., to a localization in energy space.
\end{abstract}

\section{Introduction}
The search for a quantum theory of gravity is one of the main
challenges of theoretical physics. Though until now there is no
final version of a Quantum Gravity theory it is expected that one
of the consequences of such a theory is the appearance of some
kind of space--time fluctuations, of a space--time foam. These
fluctuations may be given by fluctuations of the space--time
metric or of the connection if independent of the metric. This may
include even changes in the topology. In the most simple version
one may think of a Minkowskian background with small metrical
fluctuations. Within a semi--classical theory of quantum gravity,
space--time fluctuations are also expected to be a consequence of
fluctuations of matter fields \cite{HuVerdauger04}.

Since space--time is the arena where all physical phenomena take
place it is clear that all phenomena will be influenced by a
fluctuating space--time metric. At the first place, the
propagation of light will be influenced by a fluctuating metric
and will lead to fluctuating light cones and to a blurring of
light signals, e.g., to angular and redshift blurring
\cite{Ford05}. Since within General Relativity lengths are defined
through the time-of--flight of propagating light it is also clear
that a fluctuating metric defines a fundamental length scale and
will lead to a bound on the sharpness of length measurements
\cite{JaekelReynaud94,Klinkhamer07} which also will add additional
fundamental noise into gravitational wave detectors
\cite{Amelino-Camelia00}. This already initiated an experimental
search for a fundamental metrical noise in optical cavities
\cite{Schilleretal04}. Furthermore, metrical fluctuations have
been shown to lead to a modified inertial mass in an effective
Schr\"odinger equation derived from a Klein--Gordon equation
minimally coupled to the space--time metric
\cite{GoeklueLaemmerzahl08}. Since fluctuations in space and time
also will lead effectively to non--localities (in the sense of
higher order derivatives) in field equations it is also clear that
space--time fluctuations will also emerge in modified dispersion
relations as it has been discussed first in
\cite{Amelino-Cameliaetal98,BernadotteKlinkhamer07}. Finally, in a
recent paper by Wang and coworkers \cite{Wangetal08} it has been
shown that space--time fluctuations may be regarded as explanation
for the cosmological constant.

In this paper we further investigate the consequences of the model
introduced in \cite{GoeklueLaemmerzahl08}. We show that beside a
modification of the inertial mass followed by a violation of the
Weak Equivalence principle, the space--time fluctuations also will
lead to an {\it effective decoherence} of a quantum system.

Quantum gravity induced decoherence of quantum systems have been
considered in \cite{PowerPercival00} and
\cite{WangBinghamMendonca06}. Decoherence also appears in
discretized quantum gravity scenarios \cite{GPPQGdiscrete}. The
authors of \cite{PowerPercival00} regard space--time fluctuations
as incoherent conformal waves which are produced by
quantum--mechanical zero point fluctuations of a conformal field.
The nonlinear contribution of this field causes a decoherence of
quantum wavepackets which yields a lower bound for a parameter
which defines the borderline on which quantum to classical
transition of gravity takes place. Generally, the value for the
parameter is model--dependent. In \cite{WangBinghamMendonca06} the
authors generalize this approach accounting for spin--2
gravitational waves yielding a more optimistic lower bound on the
transition parameter which is well within an expected range for
low energy quantum gravity. They conclude that effects of quantum
fluctuations of space--time causing matter waves to lose coherence
are worth to be explored with high--sensitivity matter wave
interferometers.

\section{Quantum field in a fluctuating space--time metric}
In \cite{GoeklueLaemmerzahl08} we considered a Klein--Gordon field
minimally coupled to a space--time metric $g_{\mu\nu}$. We assumed
that this metric consists of a Minkowskian background
$\eta_{\mu\nu} = \hbox{diag}(- + + +)$ and a small fluctuating
part $|h_{\mu\nu}(x, t)| \ll 1$ so that $g_{\mu\nu}(x,t) =
\eta_{\mu\nu} + h_{\mu\nu}(x,t)$. We furthermore assumed that for
the average over space--time intervals $\langle h_{\mu\nu}(x,
t)\rangle = f_{\mu\nu}$, where all $f_{\mu\nu} = 0$ except
$f_{00}$ which we identify as a Newtonian potential.

Performing a non--relativistic limit of the Klein--Gordon equation
we obtained an effective Schr\"o\-din\-ger equation of the form
\begin{equation}
 \frac{d}{dt} |\psi(t)\rangle = -\frac{i}{\hbar} \left[ H_0 + H_p(t) \right]
|\psi(t)\rangle, \label{EffectiveSchroedinger}
\end{equation}
where ($i,j=1,2,3$)
\begin{equation} \label{DEFS}
H_0 = \frac{\mathbf{p}^2}{2m} - mU, \qquad H_p(t) = \frac{1}{2m}
\alpha^{ij}(t) p_i p_j, \qquad p_i = -i\hbar \partial_i \, .
\end{equation}
Here, the metrical fluctuations are encoded in the tensorial
function $\alpha^{ij}(t)$ which is the spatial average of squares
of the metrical fluctuations which in our approach are assumed to
consist of wavelengths short compared with the Compton wavelength
of the particle under consideration.

The tensorial function $\alpha^{ij}(t)$ is then split into an
average part and a fluctuating part
\begin{equation}\label{alphasplit}
\alpha^{ij}(t) = \tilde\alpha^{ij} + \gamma^{ij}(t) \quad
\text{with} \quad \langle \gamma^{ij}(t) \rangle = 0 \, ,
\end{equation}
where the average is denoted by angular brackets. In
\cite{GoeklueLaemmerzahl08} the part $\tilde\alpha^{ij}(t)$ has
been shown to lead to a redefinition of the inertial mass of the
quantum field under consideration. This would imply a breakdown of
the Weak Equivalence Principle which may reach a level of
$10^{-9}$ in terms of the E\"otv\"os parameter.

In the following we discuss the implications of the remaining
fluctuating part $\gamma^{ij}(t)$. We show that this term leads to
an effective decoherence of the quantum system.

\section{Derivation of the quantum master equation}\label{SEC-QMEQ}

\subsection{General form of the master equation}
Having redefined the inertial mass of the particle as described in
\cite{GoeklueLaemmerzahl08} we are left with an effective
Schr\"odinger equation of the form \eqref{EffectiveSchroedinger}
where, however, only the fluctuating part $\gamma^{ij}(t)$ enters
the Hamiltonian $H_p(t)$,
\begin{equation}
 H_p(t) = \frac{1}{2m} \gamma^{ij}(t) p_i p_j,
\end{equation}
while $H_0$ is defined as in Eq.~\eqref{DEFS} with an
appropriately renormalized inertial mass in the kinetic term. We
start by transforming to the interaction picture,
\begin{equation}
 |\psi(t)\rangle = e^{-iH_0t/\hbar} |\tilde{\psi}(t)\rangle,
\end{equation}
to obtain the Schr\"odinger equation
\begin{equation}
 \frac{d}{dt} |\tilde{\psi}(t)\rangle = -\frac{i}{\hbar} \tilde{H}_p(t)
 |\tilde{\psi}(t)\rangle, \label{SSE}
\end{equation}
where the interaction picture Hamiltonian is given by
\begin{equation}
 \tilde{H}_p(t) = e^{iH_0t/\hbar} H_p(t) e^{-iH_0t/\hbar}.
\end{equation}

Formally, Eq.~\eqref{SSE} can be regarded as a stochastic
Schr\"odinger equation (SSE) involving a random Hamiltonian
$\tilde{H}_p(t)$ with zero average, $\langle \tilde{H}_p(t)
\rangle = 0$. For a given realization of the random process
$\gamma^{ij}(t)$ the corresponding solution of the SSE represents
a pure state with the density matrix
\begin{equation}
 \tilde{R}(t) = |\tilde{\psi}(t)\rangle\langle\tilde{\psi}(t)|,
\end{equation}
satisfying the von Neumann equation
\begin{equation} \label{StochNeumann}
 \frac{d}{dt}\tilde{R}(t) = -\frac{i}{\hbar}
 \left[\tilde{H}_p(t),\tilde{R}(t)\right] \equiv
 {\mathcal{L}}(t)\tilde{R}(t),
\end{equation}
where ${\mathcal{L}}(t)$ denotes the Liouville superoperator.
However, if we consider the average over the fluctuations of the
$\gamma^{ij}(t)$ the resulting density matrix of the Schr\"odinger
particle,
\begin{equation} \label{DENSITY}
 \tilde{\rho}(t) = \left\langle \tilde{R}(t) \right\rangle =
 \left\langle |\tilde{\psi}(t)\rangle\langle\tilde{\psi}(t)| \right\rangle,
\end{equation}
generally represents a mixed quantum state. Thus, when considering
averages, the dynamics given by the SSE transforms pure states
into mixtures and leads to dissipation and decoherence processes,
i.~e., a loss of quantum coherence. Consequently, the
time-evolution of $\tilde{\rho}(t)$ is no longer given by a
unitary transformation, but must be described through a
dissipative quantum dynamical map that preserves the Hermiticity,
the trace and the positivity of the density matrix \cite{OUP}.

An efficient way of describing a quantum dynamical map consists in
the formulation of an appropriate master equation for the density
matrix $\tilde{\rho}(t)$. Thus, our goal is to derive from the
linear stochastic differential equation \eqref{StochNeumann} an
equation of motion for the average given by Eq.~\eqref{DENSITY}. A
standard approach to this problem is provided the cumulant
expansion method in which the equation of motion for
$\tilde{\rho}(t)$ is represented by means of an expansion in terms
of the ordered cumulants of the Liouville superoperator
${\mathcal{L}}(t)$ \cite{KAMPEN12}. This method is widely used in
the treatment of stochastic differential equations and in the
theory of open quantum systems \cite{OUP}. To second order in the
strength of the fluctuations it yields the equation of motion
\begin{equation} \label{MASTEREQ}
 \frac{d}{dt}\tilde{\rho}(t) = \int_0^t dt_1 \langle
 {\mathcal{L}}(t) {\mathcal{L}}(t_1) \rangle \tilde{\rho}(t)
 = -\frac{1}{\hbar^2} \int_0^t dt_1
 \left\langle \left[\tilde{H}_p(t),\left[
 \tilde{H}_p(t_1),\tilde{\rho}(t)\right]\right]\right\rangle.
\end{equation}
This is the desired quantum master equation for the density matrix
of the Schr\"odinger particle, representing a local first-order
differential equation with time-dependent coefficients.

\subsection{White noise limit and Markovian master equation}

To proceed further we have to specify the stochastic properties of
the random quantities $\gamma^{ij}(t)$. We take the simplest
ansatz assuming that the fluctuations are isotropic,
\begin{equation} \label{ansatz}
\gamma^{ij}(t) = \sigma\delta_{ij} \xi(t).
\end{equation}
It should be noted that this assumption singles out a certain
reference frame, which can be identified with the frame in which
the space averaging of Ref.~\cite{GoeklueLaemmerzahl08} has been
carried out. In Eq.~\eqref{ansatz} the function $\xi(t)$ is taken
to be a Gaussian white noise process with zero mean and a
$\delta$-shaped auto-correlation function,
\begin{equation} \label{CORR}
\langle \xi(t) \rangle = 0, \qquad \langle \xi(t) \xi(t') \rangle
= \delta(t-t').
\end{equation}
Therefore, the quantity $\sigma^2$ has the dimension of time and
we set
\begin{equation} \label{tau-c}
 \sigma^2 = \tau_c.
\end{equation}
The assumption of a white noise process means that the
auto-correlation time of the metric fluctuations is small compared
to the time scale of the free motion of the Schr\"odinger
particle. The fluctuations thus appear as un-correlated on the
time scale of the particle motion with a constant power spectrum.
In the case of colored noise with a structured power spectrum one
can determine systematic corrections to the above master equation
by means of the cumulant expansion, which generally leads to a
non-Markovian quantum master equation \cite{CPS}.

Within the white noise limit the contributions from the
higher-order cumulants vanish and the second-order master equation
\eqref{MASTEREQ} becomes an exact equation \cite{KAMPEN3}. Using
Eq.~\eqref{ansatz} we find
\begin{equation}
 \tilde{H}_p(t) = \hbar \tilde{V}(t) \xi(t),
\end{equation}
where
\begin{equation}
 \tilde{V}(t) = e^{iH_0t/\hbar} V e^{-iH_0t/\hbar}, \qquad
 V = \frac{\sqrt{\tau_c}}{\hbar} \frac{\textbf{p}^2}{2m}.
\end{equation}
Substitution into the master equation \eqref{MASTEREQ} yields
\begin{equation}
 \frac{d}{dt}\tilde{\rho}(t) = -\int_0^t dt_1
 \langle \xi(t)\xi(t_1) \rangle
 \left[\tilde{V}(t),\left[\tilde{V}(t_1),\tilde{\rho}(t)\right]\right].
\end{equation}
Employing Eq.~\eqref{CORR} we therefore get
\begin{equation}
 \frac{d}{dt}\tilde{\rho}(t) = -\frac{1}{2}
 \left[\tilde{V}(t),\left[\tilde{V}(t),\tilde{\rho}(t)\right]\right].
\end{equation}
Transforming back to the Schr\"odinger picture by means of
\begin{equation}
 \rho(t) = e^{-iH_0t/\hbar} \tilde{\rho}(t) e^{iH_0t/\hbar}
\end{equation}
we finally arrive at the master equation
\begin{equation} \label{QMEQ}
 \frac{d}{dt}\rho(t) = -\frac{i}{\hbar}[H_0,\rho(t)] +
 {\mathcal{D}}(\rho(t)),
\end{equation}
where
\begin{equation} \label{DISS}
 {\mathcal{D}}(\rho(t)) = -\frac{1}{2}
 \left[V,\left[V,\rho(t)\right]\right]
 = V\rho(t)V-\frac{1}{2}\left\{V^2,\rho(t)\right\}.
\end{equation}
Equation \eqref{QMEQ} represents a Markovian quantum master
equation for the Schr\"odinger particle. The commutator term
involving the free Hamiltonian $H_0$ describes the contribution
from the coherent motion, while the superoperator
${\mathcal{D}}(\rho)$, known as \textit{dissipator}, models all
dissipative effects. We observe that the master equation is in
Lindblad form and, thus, generates a completely positive quantum
dynamical semigroup \cite{GORINI,LINDBLAD}. We remark further that
the structure of the master equation \eqref{QMEQ} corresponds to
the so-called singular coupling limit. Within a microscopic
approach such master equations arise from the coupling of an open
quantum system to a free quantum field \cite{FRIGERIO}.

\section{Physical Implications}

\subsection{Moment equations and increase of entropy}

To discuss the physical implications of the master equation
\eqref{QMEQ} we first investigate the dynamical behavior of the
averages. The average of an arbitrary system observable $A$ is
defined by
\begin{equation}
 \langle A \rangle_t = {\mathrm{tr}} \{ A \rho(t) \},
\end{equation}
and the master equation \eqref{QMEQ} leads to the equation of
motion
\begin{equation} \label{MOMENTEQ}
 \frac{d}{dt}\langle A \rangle_t = \frac{i}{\hbar}\langle
 [H_0,A] \rangle_t - \frac{1}{2} \langle [V,[V,A]] \rangle_t.
\end{equation}
Let us consider for simplicity the case of a vanishing Newtonian
potential such that $H_0$ commutes with $V$. An immediate
consequence of Eq.~\eqref{MOMENTEQ} is then
\begin{equation} \label{MEAN-ENERGY}
 \frac{d}{dt}\langle H_0 \rangle_t = 0.
\end{equation}
This is an important property which shows that on average the
particle neither gains nor loses energy from the fluctuating
field, by contrast to other master equations proposed in the
literature. Moreover, the equations of motion for the first
moments of momentum $p_i$ and position $x_i$ are found to coincide
with those of a free particle,
\begin{equation}
 \frac{d}{dt}\langle p_i \rangle_t = 0, \qquad
 \frac{d}{dt}\langle x_i \rangle_t = \frac{1}{m} \langle p_i \rangle_t.
\end{equation}
The influence of the dissipator can however be seen in the
dynamics of the second moments. Defining the spatial variance
\begin{eqnarray}
 \sigma^2_x(t) = \langle x_i^2 \rangle_t - \langle x_i \rangle^2_t
\end{eqnarray}
we find with the help of the master equation
\begin{eqnarray} \label{x-variance}
 \sigma^2_x(t) = \sigma^2_x(0) + \frac{\sigma_{px}(0)}{m} t
 + \frac{\sigma^2_p}{m^2}t^2 + \frac{\sigma_p^2}{m^2}\tau_c t.
\end{eqnarray}
Here, the momentum variance $\sigma^2_p(t)=\langle p_i^2 \rangle_t
- \langle p_i \rangle^2_t$ is of course constant in time, and we
have evaluated the above expression in the rest frame of the
particle, assuming $\langle p_i \rangle = 0$. Moreover, we have
defined the cross-correlation $\sigma_{px}(t) = \langle p_ix_i +
x_ip_i\rangle_t - 2\langle p_i \rangle_t \langle x_i \rangle_t$.
The first three terms on the right-hand side of
Eq.~\eqref{x-variance} coincide with the corresponding expression
that is obtained from the free Schr\"odinger equation. Thus,
dissipative effects are described by the last term of
Eq.~\eqref{x-variance}. However, for large times $t\gg \tau_c$
this term is small compared to the quadratically increasing
ballistic term. Thus we see that the influence of dissipative
effects on the spreading of the wave packet is very small in the
long-time limit.

The irreversible character of the dynamics can be quantified with
the help of the dynamics of the entropy of the state $\rho(t)$.
For technical simplicity we consider here the linear entropy which
is defined by
\begin{equation} \label{entropy}
 S(t) = {\mathrm{tr}}\left\{ \rho(t) - \rho^2(t) \right\},
\end{equation}
i.~e., by one minus the purity ${\mathrm{tr}}\rho^2(t)$.
Differentiating Eq.~\eqref{entropy} and employing the master
equation \eqref{QMEQ} we obtain
\begin{equation}
 \frac{d}{dt}S(t) = {\mathrm{tr}} \left\{ W^{\dagger}(t)W(t)
 \right\} \geq 0, \qquad W(t) = [V,\rho(t)].
\end{equation}
Hence, the entropy increases monotonically because
$W^{\dagger}(t)W(t)$ is a positive operator. We also conclude from
this equation that $\dot{S}(t)=0$ if and only if $\rho(t)$
commutes with $V$, which means that $\rho(t)$ represents an
incoherent mixture of eigenstates of $V$. Since $V$ is
proportional to the kinetic energy $\rho(t)$ must be a mixture of
eigenstates of the kinetic energy, e.~g. plane or spherical waves.
In particular, a kinetic energy eigenstate is not affected by the
dissipative term and behaves exactly as for the free Schr\"odinger
equation.

\subsection{Estimation of decoherence times}
The quantum master equation \eqref{QMEQ} can easily be solved in
the momentum representation. To this end, we define the density
matrix in the momentum representation by
\begin{equation}
 \rho({\mathbf{p}},{\mathbf{p}}',t) = \langle
 {\mathbf{p}}|\rho(t)| {\mathbf{p}}' \rangle.
\end{equation}
With the help of the master equation we then find
\begin{equation}
 \frac{d}{dt}\rho({\mathbf{p}},{\mathbf{p}}',t)
 = -\frac{i}{\hbar}\left[E({\mathbf{p}})-E({\mathbf{p}}')\right]
 \rho({\mathbf{p}},{\mathbf{p}}',t)
 -\frac{\tau_c}{2\hbar^2}\left[E({\mathbf{p}})-E({\mathbf{p}}')\right]^2
 \rho({\mathbf{p}},{\mathbf{p}}',t),
\end{equation}
where $E({\mathbf{p}})={\mathbf{p}}^2/2m$. This equation is
immediately solved to yield
\begin{equation}
 \rho({\mathbf{p}},{\mathbf{p}}',t) =
 \exp\left[ -\frac{i}{\hbar}\Delta E t - \frac{(\Delta E)^2\tau_c}{2\hbar^2} t \right]
 \rho({\mathbf{p}},{\mathbf{p}}',0),
\end{equation}
where $\Delta E=E({\mathbf{p}})-E({\mathbf{p}}')$. We see that the
matrix elements corresponding to
$E({\mathbf{p}})=E({\mathbf{p}}')$ stay constant in time. In
particular, the diagonals of the density matrix in the momentum
representation are constant. On the other hand, the coherences
corresponding to different energies decay exponentially with the
rate $(\Delta E)^2\tau_c/2\hbar^2$. Thus we find an associated
decoherence time $\tau_D$ which is given by
\begin{equation} \label{tau-d}
 \tau_D = \frac{2\hbar^2}{(\Delta E)^2\tau_c} =
 2\left(\frac{\hbar}{\Delta E\cdot\tau_c}\right)^2 \tau_c.
\end{equation}
Hence, the dissipator ${\mathcal{D}}(\rho)$ of the master equation
leads to a decay of the coherences of superpositions of energy
eigenstates with different energies, resulting in an effective
dynamical localization in energy space. This feature of the master
equation is due to the fact that the fluctuating quantities
$\gamma^{ij}(t)$ couple to the components of the momentum
operator.

Let us identify the time scale $\tau_c$ that characterizes the
strength of the fluctuations (see Eqs.~\eqref{ansatz} and
\eqref{tau-c}) with the Planck time $t_p$, i.~e., we set $\tau_c =
t_p = l_p/c$ with the Planck length $l_p$. The expression
\eqref{tau-d} then yields the estimate
\begin{equation}
\tau_D \approx \frac{10^{13}{\mathrm{s}}}{(\Delta
E/{\mathrm{eV}})^2}.
\end{equation}
The decoherence time depends strongly on the scale of the energy
difference $\Delta E$. For example, $\Delta E = 1{\mathrm{eV}}$
gives a decoherence time of the order of $10^{13}$ seconds, while
$\Delta E = 1{\mathrm{MeV}}$ leads to a decoherence time of the
order of $10$ seconds.

\section{Summary and discussion}

In \cite{GoeklueLaemmerzahl08} we showed that a fluctuating
space--time metric would modify the inertial mass of quantum
particles and, thus, leads to an apparent violation of the
Equivalence Principle which gave additional motivation to
performing improved atom interferometric tests of the Equivalence
Principle. Here we derived another, complementary, implication of
such space--time fluctuations, namely decoherence of quantum
systems. In the case that the space--time fluctuations are related
to the Planck scale then the decoherence time corresponding to an
energy difference of $1$eV would be of the order of 0.3 million
years, far beyond any experimental relevance. Even if the relevant
scale is given by the grand unification scale which is three
orders of magnitude smaller than the Planck scale, the
corresponding decoherence time of about three hundred years still
is too large to be detectable. However, this result does not rule
out in general the experimental detection of dephasing effects
caused by metric fluctuations if one considers, for example,
composite quantum objects whose states can be extremely sensitive
to environmental noise.

In the derivation of our result we made two specifications. First,
we made the ansatz (\ref{ansatz}) characterizing the fluctuations.
This could in principle be generalized by introducing
off--diagonal terms in $\gamma^{ij}$. However, this is already
excluded by the averaging scheme introduced in
\cite{GoeklueLaemmerzahl08} since different components of the
fluctuating metric have been assumed to be independent of each
other. Therefore, the quadratic fluctuating tensorial quantity
$\gamma^{ij}$ defined by (\ref{alphasplit}) must be diagonal, too.
It is still possible to have different diagonal elements
describing anisotropic fluctuations leading to a replacement of
(\ref{ansatz}) by $\gamma^{ij} = \sigma_i \delta_{ij} \xi_i$.
However, since one expects that different diagonal elements of
$\gamma^{ij}$ will differ only by a tiny amount (as one expects
deviations from isotropy of space being minuscule), this should
not modify the expression for the decoherence time $\tau_D$
significantly.

Second, the white--noise scenario characterized by the two moments
(\ref{CORR}) could be generalized to colored noise. This is indeed
feasible by means of the technique indicated in
Sec.~\ref{SEC-QMEQ} and would lead to additional non--Markovian
terms in the master equation (\ref{QMEQ}). We expect however that
such terms would manifest only as small corrections to the
equation for the decoherence time $\tau_D$, which do not alter the
order of magnitude. Thus, incorporating only the simple
white--noise scenario and isotropic fluctuations is sufficient for
obtaining reasonable estimates for the decoherence time.

Finally, we emphasize that the structure of our master equation
(\ref{QMEQ}) differs significantly from the master equation which
has been derived by Power and Percival \cite{PowerPercival00} and
investigated further by Wang et al. \cite{WangBinghamMendonca06}.
Both master equations are in Lindblad form in accordance with
general principles of the theory of open quantum systems, and
describe decoherence effects that yield, for instance, a reduction
of the visibility of interference fringes. However, the position
space localization of the master equation of Power and Percival
results in an unbounded increase of the average energy, the
corresponding pointer states being given by position eigenstates.
By contrast, the master equation (\ref{QMEQ}) describes
localization in energy space with energy eigenstates as pointer
states, leading to a constant mean energy (see
Eq.~(\ref{MEAN-ENERGY})). The quantum master equation derived here
thus provides an important case of a dissipative equation of
motion which does not lead to an average increase of energy and
which could serve as a prototypical example for the
phenomenological modelling of the influence of metric fluctuations
on quantum coherence.

\section*{Acknowledgement}

We like to thank H. Dittus for fruitful discussions. HPB
gratefully acknowledges a Fellowship of the
Hanse--Wissenschaftskolleg, Delmenhorst, EG the support by the
German Research Foundation and the Centre for Quantum Engineering
and Space-Time Research QUEST, and CL the support by the German
Aerospace Center (DLR) grant no. 50WM0534.

\end{document}